\documentclass[%
 reprint,
 amsmath,amssymb,
 aps,
]{revtex4-2}
\usepackage{subfigure,amsmath,bbm}
\usepackage{color}
\usepackage{graphicx}
\usepackage{dcolumn}
\usepackage{bm}
\usepackage[ruled,vlined]{algorithm2e}

\begin{document}

\preprint{APS/123-QED}

\title{Approaching the Fundamental Limit of Orbital Angular Momentum Multiplexing Through a Hologram Metasurface}
\author{Shuai S. A. Yuan\textsuperscript{1}, Jie Wu\textsuperscript{2}, Menglin L. N. Chen\textsuperscript{3}, Zhihao Lan\textsuperscript{4}, Liang Zhang\textsuperscript{5}, Sheng Sun\textsuperscript{6}, Zhixiang Huang\textsuperscript{2}, Xiaoming Chen\textsuperscript{7}, Shilie Zheng\textsuperscript{1}, Li Jun Jiang\textsuperscript{3}, Xianmin Zhang\textsuperscript{1}}
\author{Wei E. I. Sha\textsuperscript{1,}}
\email{weisha@zju.edu.cn}
\affiliation{\textsuperscript{1}College of Information Science and Electronic Engineering, Zhejiang University, Hangzhou 310027, China}
\affiliation{\textsuperscript{2}Ministry of Education and the Key Laboratory of Electromagnetic Environmental Sensing, Department of Education of Anhui Province, Anhui University, Hefei 230039, China}
\affiliation{\textsuperscript{3}Department of Electrical and Electronic Engineering, The University of Hong Kong, Hong Kong}
\affiliation{\textsuperscript{4}Department of Electronic and Electrical Engineering, University College London, Torrington Place, London WC1E 7JE, United Kingdom}
\affiliation{\textsuperscript{5}Anhui Province Key Laboratory of Simulation and Design for Electronic Information System, Hefei Normal University, Hefei 230601, China}
\affiliation{\textsuperscript{6}School of Electronic Science and Engineering, University of Electronic Science and Technology of China, Chengdu 611731, China}
\affiliation{\textsuperscript{7}School of Information and Communications Engineering, Xi'an Jiaotong University, Xi'an 710049, China}

\begin{abstract}
Establishing and approaching the fundamental limit of orbital angular momentum (OAM) multiplexing are necessary and increasingly urgent for current multiple-input multiple-output research. In this work, we elaborate the fundamental limit in terms of independent scattering channels (or the degrees of freedom of scattered fields) through angular-spectral analysis, in conjunction with a rigorous Green’s function method. The scattering-channel limit is universal for arbitrary spatial-mode multiplexing, which is launched by a planar electromagnetic device, such as antenna, metasurface, etc., with a predefined physical size. As a proof of concept, we demonstrate both theoretically and experimentally the limit by a phase-only metasurface hologram that transforms orthogonal OAM modes to plane-wave modes scattered at critically separated angular-spectral regions. Particularly, a minimax optimization algorithm is applied to suppress angular-spectrum aliasing, achieving good performances in both full-wave simulation and experimental measurement at microwave frequencies. This work offers a theoretical upper bound and corresponding approach route for engineering designs of OAM multiplexing.
\end{abstract}

\maketitle

\section{Introduction}

The growing demand for of high data (transmission) rate per unit bandwidth has driven researchers to continuously explore the potential of independent scattering channels. It is well known that upper bound of the data rate is determined by Shannon's channel capacity \cite{shannon}; and various spatial mode multiplexing (SMM) strategies have been proposed to approach the scattering channel limit for maximizing the channel capacity \cite{agrawal1998space, evangelides1992polarization}.  Orbital angular momentum (OAM) mode is an approximate solution to the free-space Helmholtz equation in cylindrical coordinates \cite{siegman1986lasers}. In contrast to extensively-used plane-wave mode, OAM mode exhibits unique electromagnetic (EM) properties including nonuniform intensity with a phase singularity, helical wavefront, strong divergence and rich orthogonal modes \cite{allen1992orbital, LG_beam, grier2003revolution}, which has been regarded as a promising solution for overcoming the issue of limited frequency channels \cite{edfors2011orbital, wang2016advances, willner2017recent}. In the past few years, high capacity communication systems based on OAM multiplexing as well as corresponding OAM generation/detection technologies have drawn great attentions in optical \cite{bozinovic2013terabit,wang2012terabit, gibson2004free, willner2012different, zhongyiguo,oam_optical_2020, miller2019waves}, EM \cite{tamburini2012encoding, yan_multiplex, ren2017line, lee2018experimental, wang2018directly, ben_allen_oam, longli, shilie_oam, PIER, kuangzhang, xu2019interference} and acoustic \cite{shi2017high, jiang2018twisted} researches.

However, OAM multiplexing provokes a discussion about its advantages over traditional multi-input multi-output (MIMO) \cite{sarkar2019mimo, tamagnone2012comment} systems and its fundamental scattering channel limit \cite{zhao_capacity, gaffoglio_capacity,  jiexu_oam_limit}. In \cite{zhao_capacity}, an intuitive limit is given according to spatial-bandwidth product method in view of the size of focusing lens and width of OAM beams. Similarly, considering the expansion of OAM beam width during wave propagation, the aperture size and propagating distance are deemed to govern the OAM scattering channel limit, where analytical results on the numbers of independent scattering channels have been derived \cite{jiexu_oam_limit}. Based on degree of freedom (DOF) of scattered fields \cite{bucci_degrees}, the relation between the scattering channel limit and EM source size was clarified by employing multipole expansion of truncated Bessel beams over a sphere surface \cite{gaffoglio_capacity}. Moreover, information theory of metasurfaces \cite{cui_metasurface_information} is proposed to infer the DOF of plane-wave multiplexing. Nonetheless, a simple and physically intuitive bound on the number of independent scattering channels and a specific EM device designed to demonstrate the fundamental scattering channel limit would bring insights into this problem.

In this work, we establish an alternative and accessible way to deduce the scattering channel limit of OAM multiplexing, meanwhile, propose an optimization algorithm to approach the limit in ideal and realistic metasurface designs. The number of independent scattering channels of OAM modes, bounded by the physical size of the generating or detecting aperture, is equivalent to that of plane-wave modes, which could be argued from a mathematical point of view: a transformation of basis, from OAM modes to plane-wave modes and vice versa, will not change the DOF (or the number of independent scattering channels) of SMM. Consequently, the fundamental scattering channel limit is generally applicable to any SMM systems. Furthermore, a minimax algorithm could be exploited to mitigate mode crosstalks to a great extent for approaching the fundamental limit of OAM multiplexing.

This paper is organized as follows. In Section \uppercase\expandafter{\romannumeral2}, we elaborate the scattering channel limit of plane-wave modes generated by a 2-D area-constrained source. In Section \uppercase\expandafter{\romannumeral3}, we discuss that OAM basis and plane-wave basis share the same scattering channel limit when they are generated by the same size source. After that, minimax optimizations for both ideal and realistic metasurfaces are discussed in Section \uppercase\expandafter{\romannumeral4}. Then, experiment is conducted to verify the performance of the realistic metasurface at microwave frequencies in Section \uppercase\expandafter{\romannumeral5}. Additionally, the connections between EM theory and information theory of a SMM system, a rigorous deduction of the scattering channel limit based on EM Green's function and the details of minimax optimization algorithm are described in the Appendixes.

\section{Scattering channel limit of plane-wave multiplexing}
The SMM relies on the independence or orthogonality of EM modes in various bases, such as plane-wave basis and OAM basis. The data rate of the SMM system has been intensively studied in information theory, exhibiting a close connection to wave basis concept in EM theory (the relation between them is discussed in the Appendix A). Briefly speaking, the data rate critically depends on the number of available orthogonal EM modes scattered by a 2-D planar EM device or generated by a 2-D equivalent source, which can be referred as the fundamental scattering channel limit of the 2-D SMM system. In this section, we first consider the plane-wave basis, in which the derivation of the scattering channel limit (equivalent to the number of independent and distinguishable plane-wave modes generated by an area-constrained source at far-field) is straightforward.  Suppose the physical size of the area-constrained 2-D source is $L_x \times L_y$, which can be regarded as a band-limited source function $\mathbf{J}(\mathbf{r}^{\prime})$. The far-field pattern of this source can be calculated with a dyadic Green's function as
\begin{equation}\label{farfield}
\begin{aligned}
\mathbf{E}_{far}(\mathbf{k})=-j \omega \mu_{0} \frac{e^{-j k_0 r}}{4 \pi r} \int_{v}&\left(\mathbf{a}_{\theta} \mathbf{a}_{\theta}+\mathbf{a}_{\varphi} \mathbf{a}_{\varphi}\right) \cdot \mathbf{J}\left(\mathbf{r}^{\prime}\right)\\ &\exp \left(j k_0 \mathbf{r}^{\prime} \cdot \mathbf{a}_{R}\right) d \mathbf{r}^{\prime},
\end{aligned}
\end{equation}
\begin{figure}[ht!]
	\centering
	\includegraphics[width=3.4in]{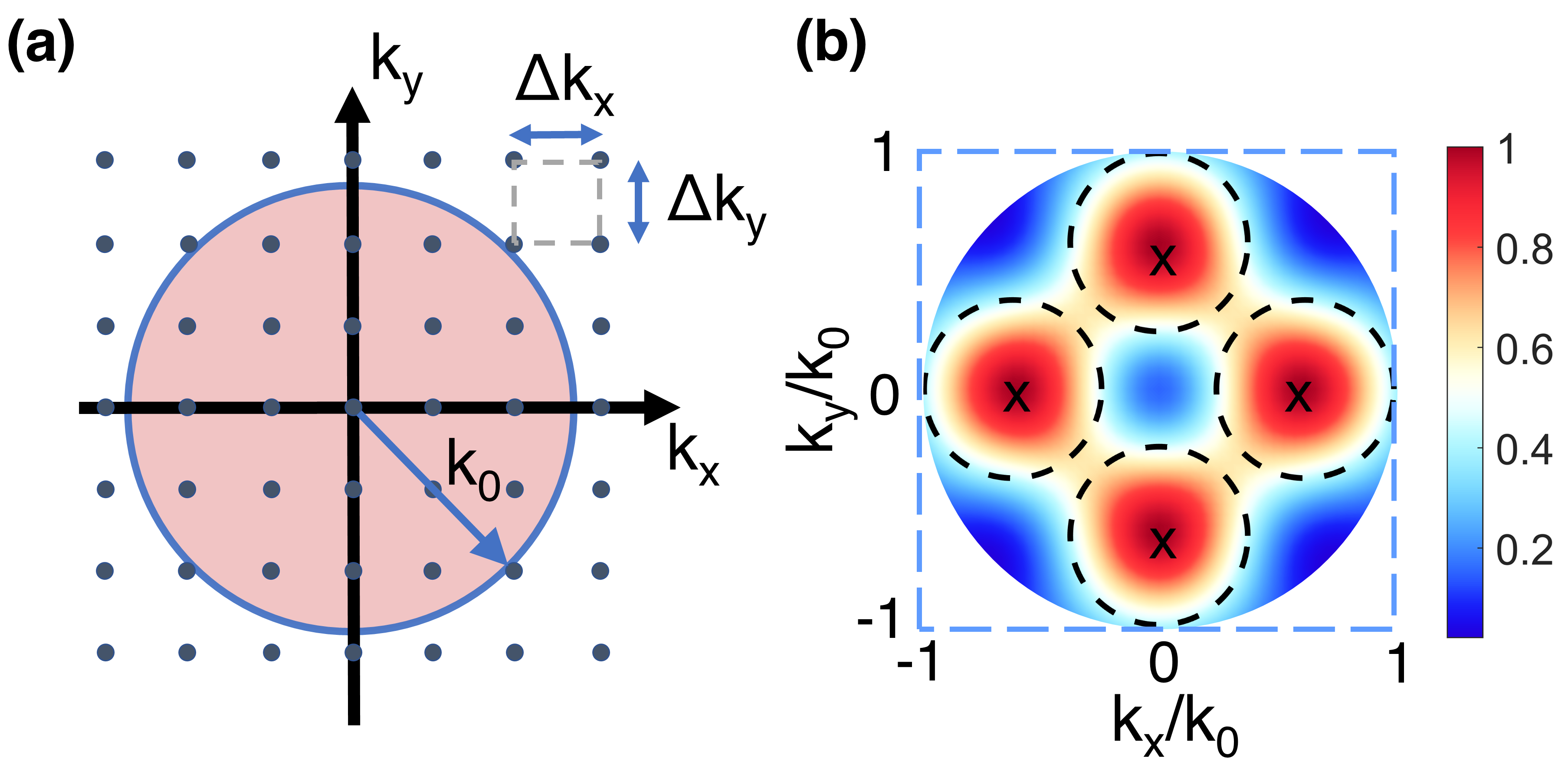}
	\caption{Fundamental scattering channel limit of plane-wave multiplexing. (a) The minimum distance between two dots represents the minimum resolution in $k$ space, the dotted square formed by adjacent four dots represents the scattering channel for a free-space propagating plane-wave mode and the light red zone inside the circle represents the total available $k$ space region. (b) Far-field pattern of multiplexed plane-wave modes generated by a square source with side length $L_x= L_y=1.12\, \lambda_0$. The radius of depicted $k$ area is $k_0$. The dotted black lines represent the half-power circles and the crosses represent the pre-designed scattering directions of the four generated plane-wave modes.}
	\label{k_domain_limit}
\end{figure}
\noindent where $\mathbf{a}_{\theta}\mathbf{a}_{\theta}$ and $\mathbf{a}_{\varphi}\mathbf{a}_{\varphi}$ represent the $\theta$ and $\varphi$ components of unit dyadic, $\mathbf{a}_{R}$ denotes the $R$ component of unit vector, both in the spherical coordinates, and $k_0$ is the wave number in free space.

 Obviously, the transform from near-field to far-field is a low-pass Fourier transform from the spatial domain to the angular-spectral domain. For plane-wave modes propagating in free space, each mode will occupy a certain $k$ area because the source area is finite or constrained. They will approach the distinguishable limit when the corresponding half-power circles of far-field patterns in the $k$ space are nearly tangent, referred to as Rayleigh's limit. Hence, on account of the limited angular-spectral region in half space, the number of distinguishable plane-wave modes as independent scattering channels can be easily deduced, inspired by the same concept in \cite{fan_fundamental_limit}.

The quantitative analysis is demonstrated as following. The minimum resolutions along the $x$ and $y$ directions in the angular-spectral domain are
\begin{equation}\label{resolution_in_k}
\Delta k_x=\frac{2\pi}{L_x},\quad \Delta k_y=\frac{2\pi}{L_y},
\end{equation}
where $L_x$ and $L_y$ are the side lengths of the source. Thus, the minimum $k$ elementary area needed for one plane-wave mode propagating in free space is $\Delta k_x\cdot\Delta k_y$, denoted by the dotted-line pixel in FIG.~\ref{k_domain_limit} (a). In free space, we have
\begin{equation}\label{k_in_free_space}
k_x ^2+k_y ^2+k_z ^2=k_0^2,
\end{equation}
where $k_0$ is the free space wave number. For a propagating plane-wave mode which obeys $k_z^2>0$, the total angular-spectral area in the half space is bounded by
\begin{equation}\label{k_resource_of_half_space}
k_x ^2+k_y ^2<k_0^2,
\end{equation}
marked by the light red circle in FIG.~\ref{k_domain_limit} (a). Then, the number of independent scattering channels can be obtained as
\begin{equation}\label{upper_limit}
N \leq   \frac{\pi k_0^2}{\Delta k_x\cdot \Delta k_y}=\frac{\pi(\frac{2\pi}{\lambda_0})^2}{\frac{2\pi}{L_x}\cdot\frac{2\pi}{L_y}}=\frac{\pi S}{\lambda_0^2},
\end{equation}
where $N$ is the fundamental limit of independent scattering channels of plane-wave multiplexing, $S=L_x\cdot L_y$ is the geometric area of the source and $\lambda_0$ is the free space wavelength. This limit considers only one type of polarization, and it should be doubled if dual polarizations are adopted. If more plane-wave modes beyond this limit are added, they will not be distinguishable or angularly resolved. Evanescent modes are required to support the expanded angular spectrum, which, however, are not suitable for far-field communication. Moreover, it is worth noting that the Eq. (5) is rather an intuitive method, we also provide a strict and rigorous method to deduce this limit based on EM Green's function in Appendix B, which yields almost the same result.

Here, an example is given for validation. The far-field pattern of four multiplexed plane-wave modes oriented at four pre-designed directions is depicted in FIG.~\ref{k_domain_limit} (b). The minimum side length of the source is set as $L_x= L_y=1.12\, \lambda_0$ according to Eq. \eqref{upper_limit}. The far-field is calculated by
\begin{equation}\label{plane wave test}
P_{far}={\left| \sum_{n=1}^4 \mathcal{F}\bigg\{\exp[jk_x(n)\cdot x+jk_y(n)\cdot y]\cdot \Omega(x,y)\bigg\}\right|}^2,
\end{equation}
where $\mathcal{F}$ denotes Fourier transform, $\Omega(x,y)$ is a window function with $\Omega=1$ ($\Omega=0$) inside (outside) the source area, $n$ is the index of plane-wave modes, and the corresponding wave numbers of the four pre-designed directions are $k_{x}(n)/k_0=[0.65,0,-0.65,0]$ and $k_{y}(n)/k_0=[0,0.65,0,-0.65]$. It can be visually observed from FIG.~\ref{k_domain_limit} (b) that the half-power circle of each mode is nearly tangent to that of nearby modes. Also, the $k$ space is almost totally occupied with little wide-angle margin, validating Eq. \eqref{upper_limit}. Quantitatively, if we define the crosstalk between two modes as the inner product of there far-field patterns, the normalized crosstalk matrix is
\begin{equation}
	\mathbf{A_p}=\left[\begin{array}{cccc}
	1&0.06&0.06&0.06 \\
	0.06& 1& 0.06& 0.06 \\
	0.06& 0.06& 1& 0.06\\
	0.06& 0.06& 0.06& 1\\
	\end{array}\right],
\end{equation}
where the element $\mathbf{A}_{m,n}$ denotes the crosstalk between mode $m$ and mode $n$, and $m, n\in$ [1, 2, 3, 4].
\section{The equivalence of scattering channel limit between OAM basis and plane-wave basis}
\begin{figure}[ht!]
	\centering
	\includegraphics[width=3.4in]{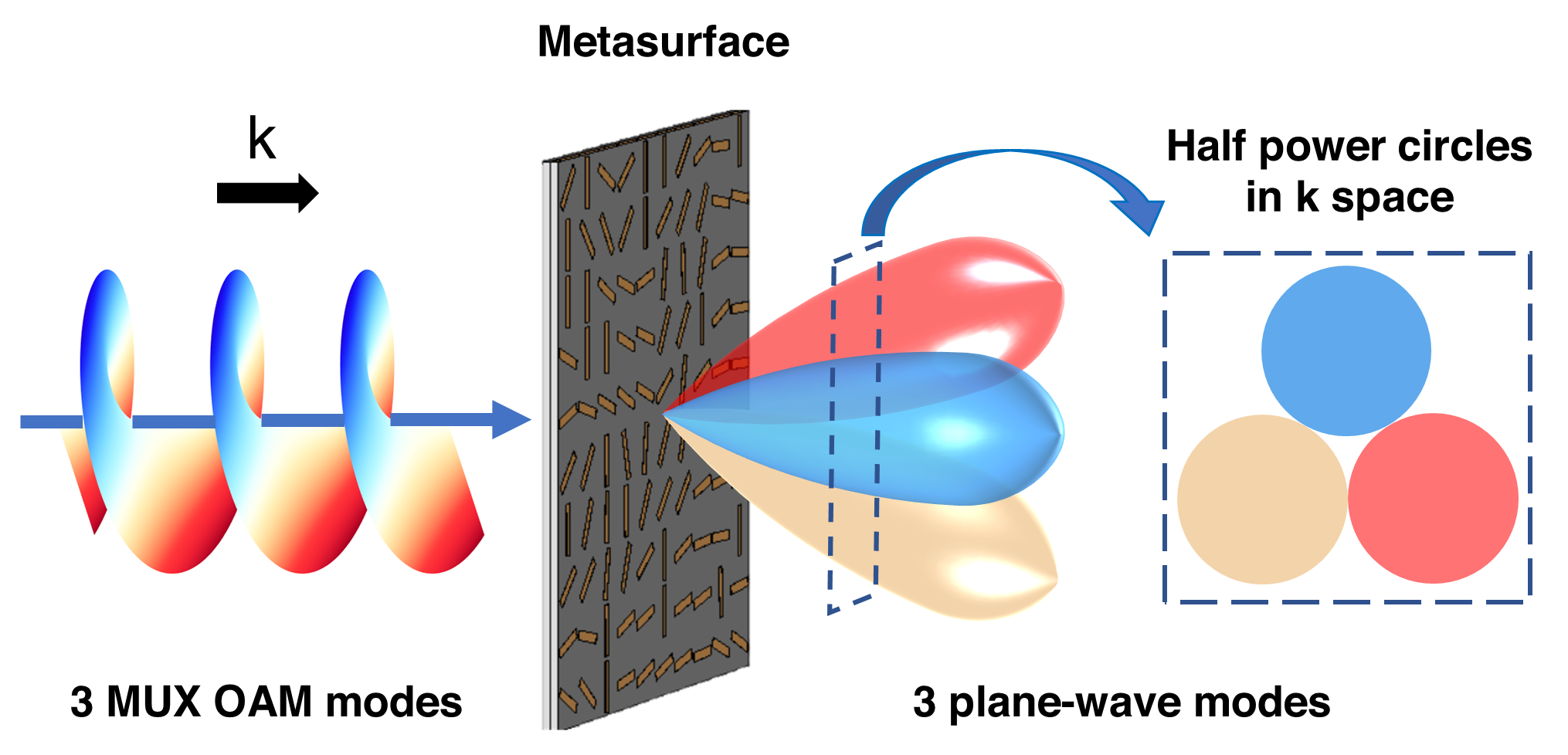}
	\caption{Schematic diagram to illustrate the fundamental scattering channel limit of OAM multiplexing based on the transformation of basis. The mixed OAM wave, generated by a superposition of three multiplexed and orthogonal OAM modes with encoded amplitudes, is transformed to three propagating and orthogonal plane-wave modes by a detecting metasurface. When the number of superposed OAM modes approaches the fundamental limit determined by the size of the metasurface, the half-power circle of each plane-wave mode will be tangent to that of nearby modes.}
	\label{oam_communication_system}
\end{figure}
Different from plane-wave basis, spatial waves or structured waves, such as Laguerre-Gaussian beams carrying OAM, form the spatially orthogonal basis which can be multiplexed towards unidirection \cite{LG_beam}. Information is encoded (generated) or decoded (detected) in terms of the superposed orthogonal modes via a EM generator or detector involving antenna arrays, metasurfaces, spiral phase plates, etc. Particularly, the EM detector is capable of distinguishing these modes for realizing information retrieval. Intuitively, the structured waves can be infinitely multiplexed in a line of sight (LOS) communication system. However, the maximum number of orthogonal spatial modes is only limited by the physical or geometrical size of the EM device, and is independent of the specific source basis. Although there could be infinite orthogonal source distributions in the source plane, for example, infinite orthogonal LG modes (even with different beam waists \cite{vallone2017role, SM}) could be arranged in a truncated plane, the available number of scattering channels is still limited due to the constraint from the receiving side.

Hence, as a proof of concept, we can design a phase-only metasurface to unitarily transform multiplexed OAM modes to plane-wave modes in limit-approaching situation, benefiting from its remarkable advances in OAM multiplexing \cite{tan2019free, jiang2018highly, zhao2017demonstration, chen2018digitalizing}. As shown in FIG.~\ref{oam_communication_system}, for a set of multiplexed OAM modes with encoded amplitudes propagating towards one direction, we expect that each of them can be transformed to one plane-wave mode propagating towards a pre-designed direction with a well-engineered metasurface. As a time-reversal procedure due to reciprocity, multiplexed OAM modes can be generated by impinging the metasurface with plane-wave modes from different directions. The proposed fundamental scattering channel limit can be approached if corresponding orthogonal plane-wave modes are generated and angularly distinguishable. In this case, the information encoded in the amplitudes of multiplexed OAM modes is transmitted to that in the amplitudes of multiplexed plane-wave modes. Nevertheless, maximizing the number of independent scattering channels is difficult because the transformation of basis by single metasurface is mathematically non-ideal. Under the incidence of one OAM mode, several modes will be generated (one plane-wave mode and other OAM modes), where only the plane-wave mode is useful for information retrieval, and the other OAM modes are regarded as crosstalks. Hence, in order to approach the scattering channel limit, minimax algorithm is employed to optimize the metasurface in the next section.

\section{Approach the fundamental scattering channel limit with minimax optimization}
To move from theory to practice, in this section, both ideal and realistic metasurfaces are studied to verify the performance of minimax optimization. In order to implement an optimization algorithm, input variables and objective functions should be given. As we only consider lossless metasurfaces, input variables will be thus the phases of the discretized meta-atoms of corresponding metasurfaces. For both ideal and realistic metasurfaces, objective functions are constructed similarly by the far-field patterns of the metasurfaces under the incidences of multiplexed OAM modes. However, the methods of calculating the far-field patterns with the input variables are different for the ideal and realistic cases. Technical details and results are discussed as following.
\subsection{Ideal Metasurface}
To begin with, we consider the minimum source size for spatial multiplexing of four OAM modes, then extend them to eight modes. In order to match the full-wave EM simulation in the next subsection, the metasurface is designed to be composed of $10\times10$ meta-atoms working at 5.088\,GHz, in which the meta-atoms refer to the repetitive unit cells of the metasurface. The area of this metasurface is set to be $S=1.36\, \lambda_0^2$ and the side length is $L_x=L_y=1.17\lambda_0$. From Eq. \eqref{upper_limit}, the maximum number of independent OAM modes allowed for multiplexing is calculated to be $4.3$, leaving negligible margin for the multiplexing of four modes. According to the method proposed in \cite{menglin_detection}, multiplexed OAM modes can be detected by a single metasurface with phase distribution
\begin{equation}\label{transmittance_of_meta}
\begin{aligned}
t(r, \phi)=A(r, \phi)\exp\left(j\phi_0(r, \phi)\right)\cdot \\
\left\{\sum_{n} \exp\left[j\left(l_{n} \phi+k_{x }(n) x+k_{y }(n) y\right)\right]\right\},
\end{aligned}
\end{equation}
where $n$ is the index of modes, $l_n$ is the topological charge of OAM modes, $A(r, \phi)$ are the normalized amplitude factors for guaranteeing $|t(r, \phi)|=1$ and $\phi_0(r, \phi)$ are the initial phases to be optimized for reducing mode crosstalks of high-order OAM modes and suppressing side lobes of transformed plane-wave modes. Laguerre Gaussian (LG) beams are utilized as OAM sources with the radical index $p = 0$ and azimuthal index $l_0$ at the position $z=0$ \cite{LG_beam}
\begin{equation}\label{LG_beam}
\begin{aligned}
L G_{0 \l_0}= \sqrt{\frac{2}{\pi(|l_0|) !}} \frac{1}{w(0)}\left[\frac{r \sqrt{2}}{w(0)}\right]^{|l_0|} \exp &\left[\frac{-r^{2}}{w^{2}(0)}\right] \exp [j l_0 \phi] \\
& \cdot L_{0}^{|l_0|}\left(\frac{2 r^{2}}{w^{2}(0)}\right),
\end{aligned}
\end{equation}
where $r$ is the radius, $\phi$ is the azimuthal angle, $L_0$ is the associated Laguerre polynomial and $w(0)$ is the beam waist. It is worth noting that $w(0)$ is set equal to the side length of the metasurface. Next, a phase plane of the same size is digitized into $10\times10$ pixels to represent the ideal metasurface and additional phase will be carried when the LG beams pass through the plane.  After that, the far-field pattern can be calculated with a Fourier transform \cite{goodman2005introduction}.
\begin{equation}\label{LG_farfield}
E_{far}=\mathcal{F}\bigg\{L G_{0 \l_0}\cdot t(r,\phi)\bigg\}
=\sum_{n}  E_{OAM(l_n+l_0)}(k_x,k_y),
\end{equation}
where $\mathcal{F}$ denotes Fourier transform, $t(r,\phi)$ is the phase distribution of the metasurface and $(l_n+l_0)$ represents the orders of generated OAM modes. The normalized far-field intensity pattern can be written as
\begin{equation}\label{normalized_power}
P_{far}(k_x,k_y)=\frac{E_{far}^2(k_x,k_y)}{\sum_{k_x,k_y} E_{far}^2(k_x,k_y)},
\end{equation}
where $\sum_{k_x,k_y}$ denotes the sum of power density over all $k$ space. Obviously, far-field intensity $P_{far}$ is a function of $k_x$ and $k_y$, denoting the power intensities along different directions. By engineering the phase distribution of the metasurface, the OAM mode with $l_0+l_n=0$, which has the highest intensity and behaves like a plane wave mode, can be generated towards a pre-designed direction to realize the mode detection and information retrieval.

Unfortunately, the unwanted high-order OAM modes with $l_0+l_n\neq0$ are inevitable with this detecting method. As the size of metasurface becomes smaller, the unwanted crosstalks between the high-order modes become worse, especially for the minimum limiting case, which shows a big difference from our previous work \cite{menglin_detection}. Such side lobes will not only reduce the detecting accuracy but will also make it difficult to analyze the scattering channel limit. 
\begin{figure}[ht!]
	\centering
	\includegraphics[width=3.4in]{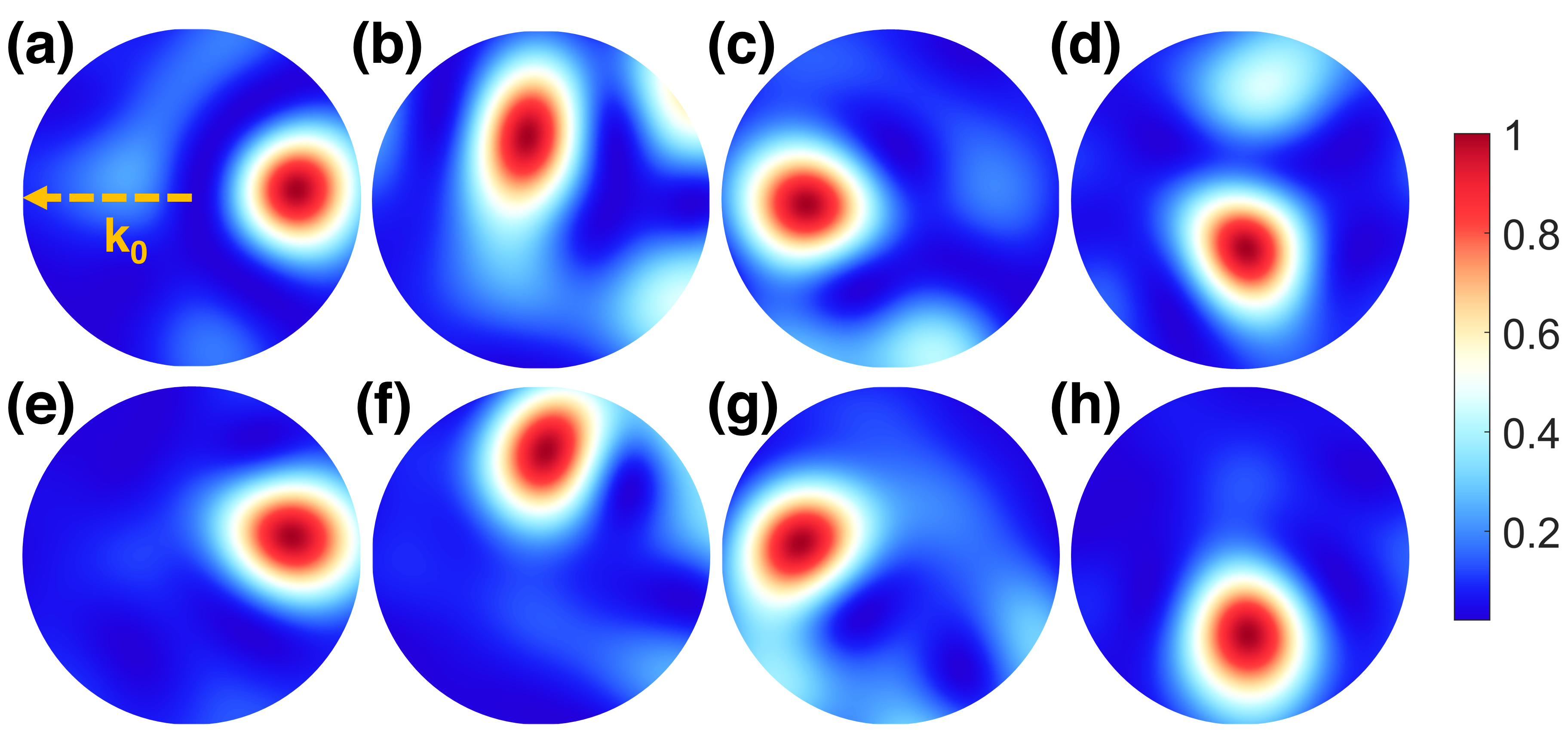}
	\caption{Responses of an ideal metasurface for OAM detecting under the incidences of different OAM modes. (a)-(d) Far-field intensity patterns under the incidences of OAM modes $l_0=1, 3, -3, -1$ without optimization; (e)-(h) Far-field intensity patterns under the incidences of OAM modes $l_0=1, 3, -3, -1$ with optimization.}
	\label{farfield_intensity_optimization}
\end{figure}
The design of this OAM sorter is then in virtue of an optimization algorithm. Mathematically, the optimization in this case can be considered as a worst-case tolerance problem, and the key is to find the phase distribution which makes the worst far-field pattern optimal under the incidence of each OAM mode to be multiplexed. Therefore, each input OAM mode is related to one dash-line circle in angular spectrum in Fig. 1 (b), the best performance will be achieved if we make the power of OAM modes diffracted into the corresponding angular circles as much as possible. We propose a minimax optimization algorithm to suppress the unwanted side lobes as much as possible, details of the optimization algorithm can be found in Appendix C, where the input variables are the phases of the discretized pixels (ideal meta-atoms) denoted by $[\phi_1,\phi_2,\cdots\phi_{100}]$, multiplexed LG beams are fixed as the input sources and objective functions are constructed with far-field patterns given by Eqs. (7-10). 

\begin{figure}[ht!]
	\centering
	\includegraphics[width=3.4in]{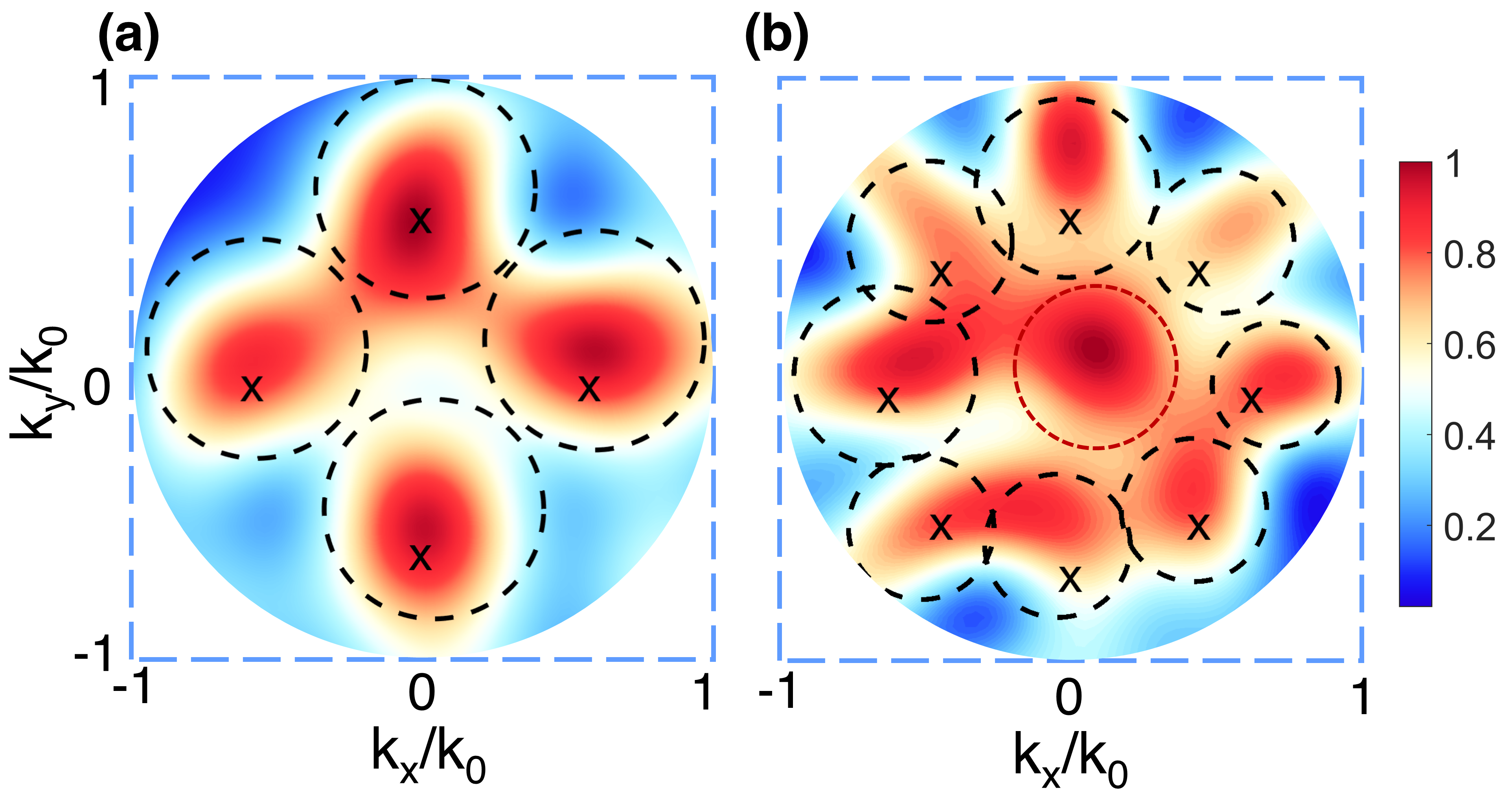}
	\caption{Responses of the ideal metasurfaces under the incidences of multiplexed OAM modes. (a) Far-field intensity pattern under the incidence of multiplexed OAM modes $l_0=1, 3, -3, -1$. The dotted black lines represent the half-power circles of far-field intensity patterns and the crosses represent the pre-designed scattered directions of the four OAM modes (the scattering channel limit of OAM multiplexing with this metasurface is 4.27); (b) Far-field intensity pattern under the incidence of multiplexed OAM modes $l_0=1, 3, 5, 7, -7, -5, -3, -1$. The dotted black lines represent the half-power circles of far-field intensity patterns, the dotted red line represents the overlapped mode and the crosses represent the pre-designed scattered directions of the eight OAM modes (the scattering channel limit of OAM multiplexing with this metasurface is 8.4).}
	\label{farfield_intensity_multiplexed}
\end{figure}

The comparison between non-optimized and optimized results is depicted in FIG.~\ref{farfield_intensity_optimization}, from which one can see that the side lobes are significantly suppressed, especially for the modes 1 and -1.  With the optimization, the normalized far-field intensity patterns under the incidences of multiplexed OAM modes are shown in FIG.~\ref{farfield_intensity_multiplexed}. For each scattered plane-wave mode, the maximum-intensity point in the $k$ space with a normalized power intensity 1 is set as the center of the half-power circle, and the radius of the half-power circle is set as the median of the distance between the two points with the normalized power intensity 1 and 0.5 in the $k$ space, respectively. Furthermore, the pre-designed scattered directions of the four OAM modes are denoted by the crosses. There will be minor deviations between the designed and realistic scattered directions. For the four OAM modes multiplexing, half-power circle of each scattered plane-wave mode is nearly tangent to that of nearby modes and almost all of the $k$ space is occupied, manifesting that the OAM scattering channel limit is the same as that of the plane-wave case in Section II. Calculated with the same method as Eq. (7), the crosstalk matrix of the four generated modes is
\begin{equation}
\mathbf{A_o}=\left[\begin{array}{cccc}
1&0.18&0.22&0.10 \\
018& 1& 0.28& 0.18 \\
0.22& 0.28& 1& 0.18\\
0.10& 0.18& 0.18& 1\\
\end{array}\right],
\end{equation}
where the crosstalk between the modes $l_0$ = -3 and 3 is the worst case. Moreover, eight OAM modes multiplexing is also investigated with a larger metasurface whose size is $S=2.67 \lambda_0^2$, and the theoretical upper limit $N$ in Eq. \eqref{upper_limit} is calculated to be 8.4. In FIG.~\ref{farfield_intensity_multiplexed}(b), the red-dot circle denotes an overlapped useless field region resulting from the crosstalks of high-order OAM modes. Fortunately, eight transformed plane-wave modes are angular-spectrally separated in a marginal sense. Therefore, the fundamental scattering channel limit of OAM multiplexing can be roughly approached with the proposed optimization algorithm.
\subsection{Realistic Metasurface}
In this subsection, a realistic metasurface is built up, where the input variables, sources, calculation of far-field and error compensations are different from the ideal case. The area of this realistic metasurface is set to be $S=1.36\, \lambda_0^2$ and it is constructed by the same $10\times10$ meta-atoms as the ideal case. For flexible phase control and simple printed-circuit-board fabrication, Pancharatnam-Berry meta-atoms are chosen to construct the detecting metasurface for achieving desired geometric phase \cite{menglin_generation, ding2015ultrathin, PB_zhoulei}. Under the incidence of circular-polarized wave, additional phase will be carried by the transmitted cross-polarized wave and the introduced phase by each meta-atom is exactly twice of its rotation angle. 
\begin{figure}[ht!]
	\centering
	\includegraphics[width=3.4in]{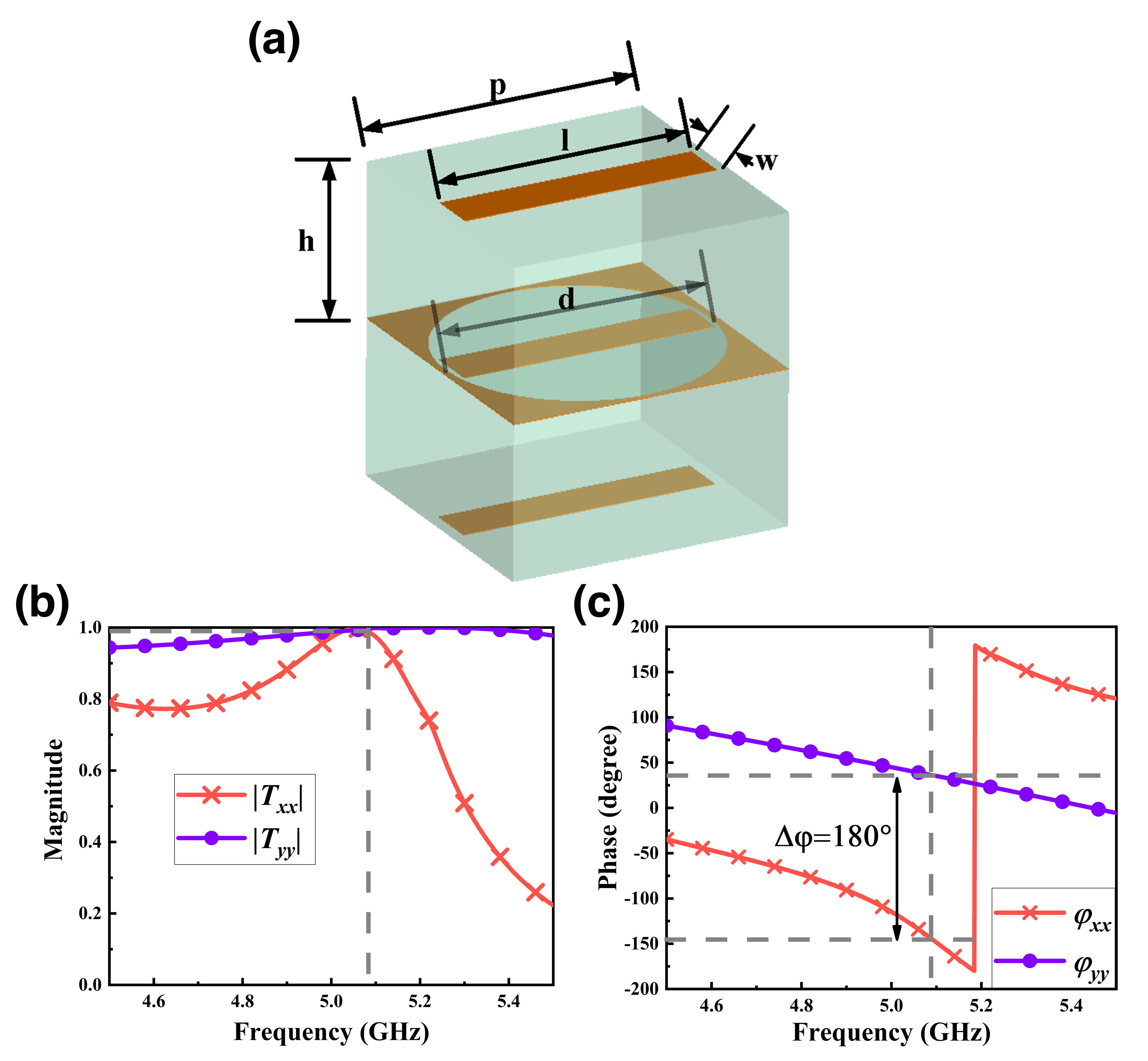}
	\caption{Full-wave simulation results of the meta-atom. (a) Structure of the meta-atom. The parameters are $\epsilon_r=12.2, p=7$ mm, $h=3.81$ mm, $l=6.4$ mm, $d=6.7$ mm and $w=1.2$ mm; (b) Simulated magnitudes of the co-transmission coefficients; (c) Simulated phases of the co-transmission coefficients.}
	\label{meta_atom_design}
\end{figure}
It is noticeable that the size of this detecting metasurface is minuscule with the side length $L_x=L_y=1.17\, \lambda_0$, which suggests small size meta-atoms are needed. Based on the design principle in \cite{PB_zhoulei}, we construct a meta-atom around $0.11\times0.11\, \lambda_0^2$ by using a substrate with a high relative dielectric constant $\epsilon_r=12.2$.  Using periodic boundary conditions in full-wave EM simulation, this meta-atom achieves nearly $99.8\,\%$ transmission efficiency at $5.088\,$GHz as shown in FIG. \ref{meta_atom_design}. And the detecting metasurface, composed of $10\times10$ meta-atoms, is shown in FIG.~\ref{metasurface_pattern}.

As we adopt the geometric phase based metasurface, all the variables to be optimized are the rotation angles of the meta-atoms represented by $[\alpha_1,\alpha_2,\cdots \alpha_{100}]$. The far-field calculation is different from that of previous case. In the ideal case, totally confined LG beams and ideal phase plane are utilized, which are not attainable in realistic implementation. For general consideration, circularly polarized plane-wave modes are used to generate four OAM modes with the help of corresponding generating metasurfaces. Different from the ideal case, hypergeometric Gaussian (HyGG) beams are generated with this method, which appears more radial-mode components and energy loss \cite{karimi2007hypergeometric,zhong2021gouy}. Also, high-order diffraction induced by pixel structure is inevitable in realistic hologram. Therefore, even with the same optimization algorithm, the performance of side lobe suppression is expected to be worse in realistic case. The response of the detecting metasurface can be obtained by superposing the transmissions of the four multiplexed OAM modes with the same input power. 
\begin{figure}[ht!]
	\centering
	\includegraphics[width=3.4in]{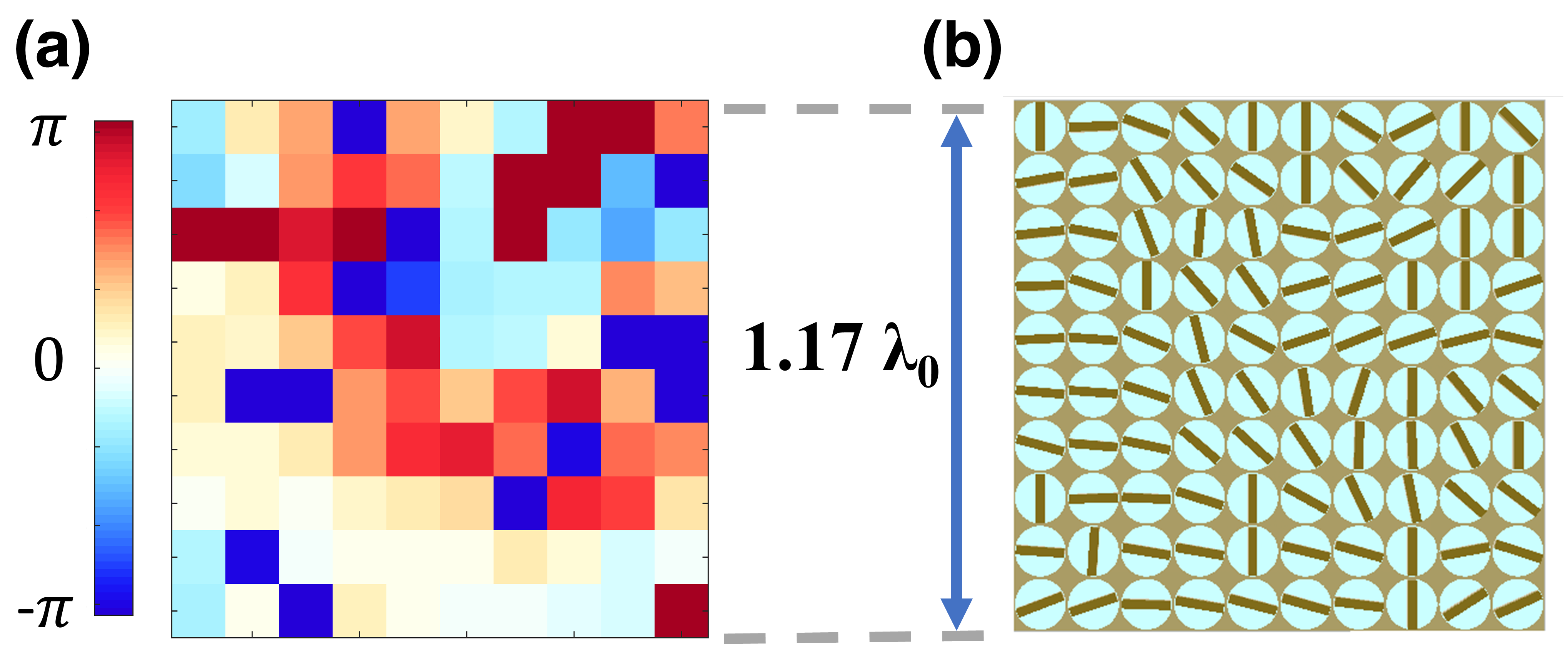}
	\caption{Optimized pattern of the detecting metasurface. (a) Discretized phase distribution of the metasurface; (b) Top view of the metasurface.}
	\label{metasurface_pattern}
\end{figure}
To detect OAM modes with better efficiency (minimized crosstalks) at microwave frequencies, topology charges of $l_0=1, 2, -2, -1$ are adopted instead of $l_0=1, 3, -3, -1$ as in the ideal metasurface case after a comparison. As for the phase distribution design, a challenging problem is that the phase introduced by metasurface is always non-ideal because of the breakdown of periodic boundary conditions after rotations of meta-atoms. Several methods are proposed to solve this problem.

(1) The phase introduced by each meta-atom could be obtained by simulating the entire metasurface rather than using periodic boundary conditions. Each meta-atom can be represented by a transmission matrix
\begin{equation}\label{T_matrix}
\mathbf{T}=
\begin{bmatrix}t_{11}&t_{12}\\t_{21}&t_{22}\end{bmatrix},
\end{equation}
and the elements of the transmission matrix can be extracted from the full-wave simulation of the entire metasurface.

(2) When the meta-atom is rotated by $\alpha$, the transmission matrix becomes
\begin{equation}\label{rotation_of_T}
\mathbf{T}(\alpha)=
\begin{bmatrix}\cos(-\alpha)&\sin(-\alpha)\\-\sin(-\alpha)&\cos(-\alpha)\end{bmatrix}\mathbf{T}\begin{bmatrix}\cos(\alpha)&\sin(\alpha)\\-\sin(\alpha)&\cos(\alpha)\end{bmatrix}.
\end{equation}
As a result, the transmission matrix of each meta-atom can be calculated by their rotation angles $[\alpha_1,\alpha_2,\cdots \alpha_{100}]$.

(3) OAM modes with $l_0=1, 2, -2, -1$ are generated by four metasurfaces with different phase distributions as excitation sources. Even if the generated OAM modes are not pure, the OAM sources $E_{i}$ can be sampled from full-wave simulation to avoid the error.

With these methods, we can reduce the errors from the breakdown of periodic boundary conditions and impure OAM sources. Nonetheless, phase error still exists, which could result from the inevitable coupling between meta-atoms.

The output near-field $E_{o}$ can be calculated by multiplying rotated transmission matrices with the $E_{i}$ fields at corresponding meta-atom centers.
\begin{figure}[ht!]
	\centering
	\includegraphics[width=3.3in]{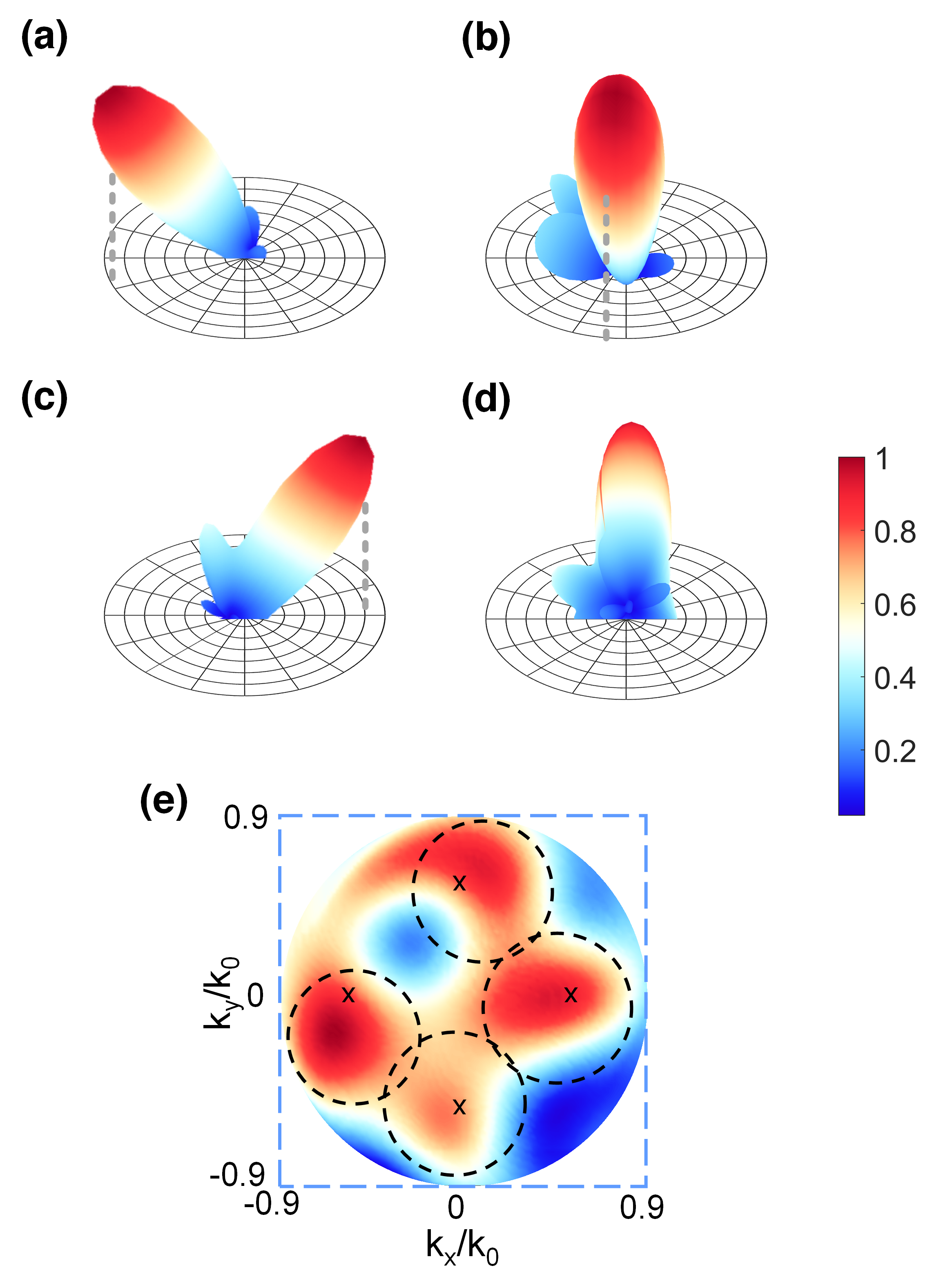}
	\caption{Full-wave EM simulations of far-field patterns. (a)-(d) Normalized 3-D patterns under the incidences of OAM modes $l_0=1, 2, -2, -1$; (e) Normalized 2-D pattern in $k$ space under the incidence of multiplexed OAM modes. The dotted black lines represent the half-power circles of far-field intensity patterns and the crosses represent the pre-designed scattered directions of the four OAM modes.}
	\label{full_wave_farfield_pattern}
\end{figure}
\begin{equation}\label{E_nearfield}
\begin{bmatrix}E^{xp}_{o}(n)\\E^{yp}_{o}(n)\end{bmatrix}=\mathbf{T}(\alpha_n)\cdot\begin{bmatrix}E^{xp}_{i}(n)\\E^{yp}_{i}(n)\end{bmatrix},
\end{equation}
where $n=1,\cdots, 100$ is the index of each meta-atom, $\alpha_n$ is the rotation angle of each meta-atom, and superscripts $xp$ and $yp$ denote the $x$-polarized and $y$-polarized components. With the output near-field, far-field can be obtained by Fresnel diffraction \cite{born2013principles}
\begin{equation}
\begin{split}
E_{far}(x, y)=\frac{e^{i k_0 z}}{i \lambda_0 z} \iint_{-\infty}^{+\infty} &E_{o}\left(x^{\prime}, y^{\prime}, 0\right) \\&\cdot e^{\frac{i k_0}{2 z}\left[\left(x-x^{\prime}\right)^{2}+\left(y-y^{\prime}\right)^{2}\right]} d x^{\prime} d y^{\prime},
\end{split}
\end{equation}
where $z$ is the distance between the source plane and target plane, and $x^{\prime}$ and $y^{\prime}$ denote the geometric positions at source plane. The left-polarized and right-polarized components are taken as
\begin{equation}\label{E_left}
E_{l}=\frac{E^{xp}_{far}+iE^{yp}_{far}}{2},\,E_{r}=\frac{E^{xp}_{far}-iE^{yp}_{far}}{2}.
\end{equation}
\begin{figure}[ht!]
	\centering
	\includegraphics[width=3.3in]{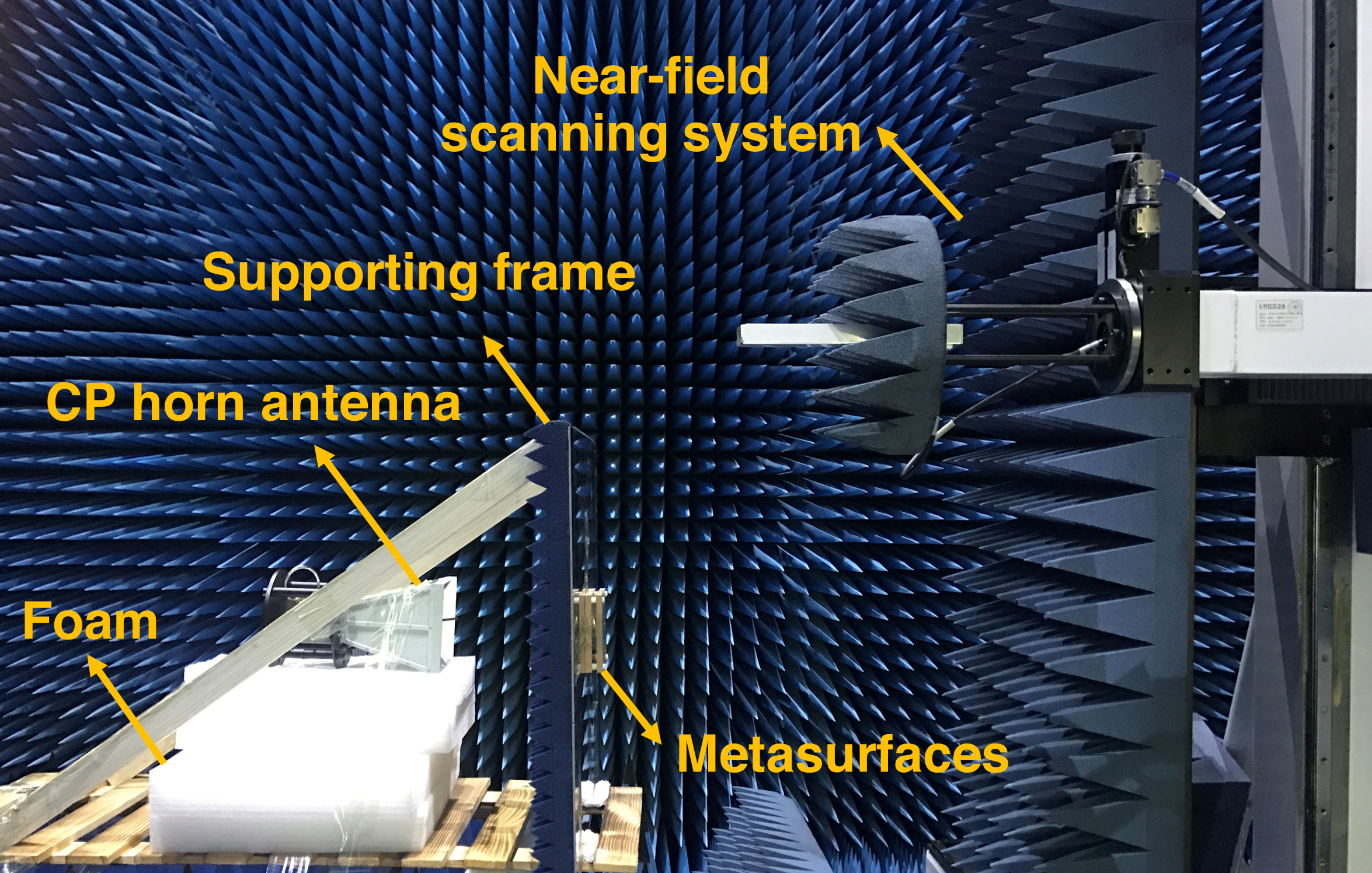}
	\caption{Experimental setup of the near-field measurement.}
	\label{experiment set up}
\end{figure}
Assuming the incident wave is left circularly-polarized, the objective polarization will be transformed twice in total (one is for the generation of OAM source and another for detection), which will end up being still left circularly-polarized. Hence, the objective far-field intensity is
\begin{equation}\label{P_right}
P_{far}(k_x,k_y)=\frac{E_{l}^2(k_x,k_y)}{\sum_{k_x,k_y} E_{l}^2(k_x,k_y)},
\end{equation}
Then, the optimization algorithm can be constructed similarly to the ideal case, as described in Appendix C, where the input variables are the rotation angles of the meta-atoms, superposed E-field of four OAM modes generated by the generating metasurfaces is employed as the input source, and the far-field patterns are calculated with Eqs. (11-16). Optimized far-field patterns are depicted in FIG.~\ref{full_wave_farfield_pattern}, where the scattered waves have relatively low side lobes. Far-field patterns without optimization can be found in Supplemental Material \cite{SM}, which shows severe side lobes. The response of the detecting metasurface under the incidences of the multiplexed OAM modes is also given as a comparison to the ideal case result in FIG. \ref{farfield_intensity_multiplexed}(a). The corresponding crosstalk matrix is
\begin{equation}
\mathbf{A_{on}}=\left[\begin{array}{cccc}
1&0.23&0.19&0.22 \\
0.23& 1& 0.17& 0.12 \\
0.19& 0.17& 1& 0.32\\
0.22& 0.12& 0.32& 1\\
\end{array}\right],
\end{equation}
where the crosstalk between the modes $l_0$ = -2 and -1 is the worst case. Although there appear some deviations of wave orientations and deformations of wave shapes due to the mode crosstalks, the four modes are generally separated. Hence, the fundamental scattering channel limit of OAM multiplexing can also be roughly approached with the realistic detecting metasurface.
\section{Microwave Experiments}
Experimental facilities and measured results are given in this section. As shown in FIG.~\ref{experiment set up}, experimental facilities mainly consist of circularly-polarized (CP) horn antenna, generating and detecting metasurfaces, supporting frame and near-field scanning system. The circularly-polarized horn antenna has an aperture size of 128$\times$128 mm, which emits left circularly-polarized wave at 5.088\,GHz.  The HyGG OAM modes carrying topology charges $l_0=1, 2, -2, -1$ are generated with corresponding four generating metasurfaces, then impinge on the optimized detecting metasurface. The distance between the two metasurfaces is fixed as 6 mm by a plastic spacer to avoid severe divergences of OAM modes, while focusing lens should be applied if a long distance communication is considered. A 500$\times$500 mm iron plate with a 70$\times$70 square hole in the center is fabricated as the supporting frame, which is covered by a piece of the same shaped absorption material to reduce the edge scattering of the metal hole. This supporting frame is used for guaranteeing that waves can only pass through the metasurface rather than bypass it. The details of fabricated metasurfaces and supporting frame are given in Supplemental Material \cite{SM}. Near-field scanning system is employed for collecting the data of near-field electric components $E^{xp}$ and $E^{yp}$, from which left circularly-polarized far-field pattern can be calculated.  As testing criteria, the distance between the metasurface and probe is set as 170 mm, and 52$\times$52 data points are sampled over a 800$\times$800 mm plane.

\begin{figure}[ht!]
	\centering
	\includegraphics[width=3.3in]{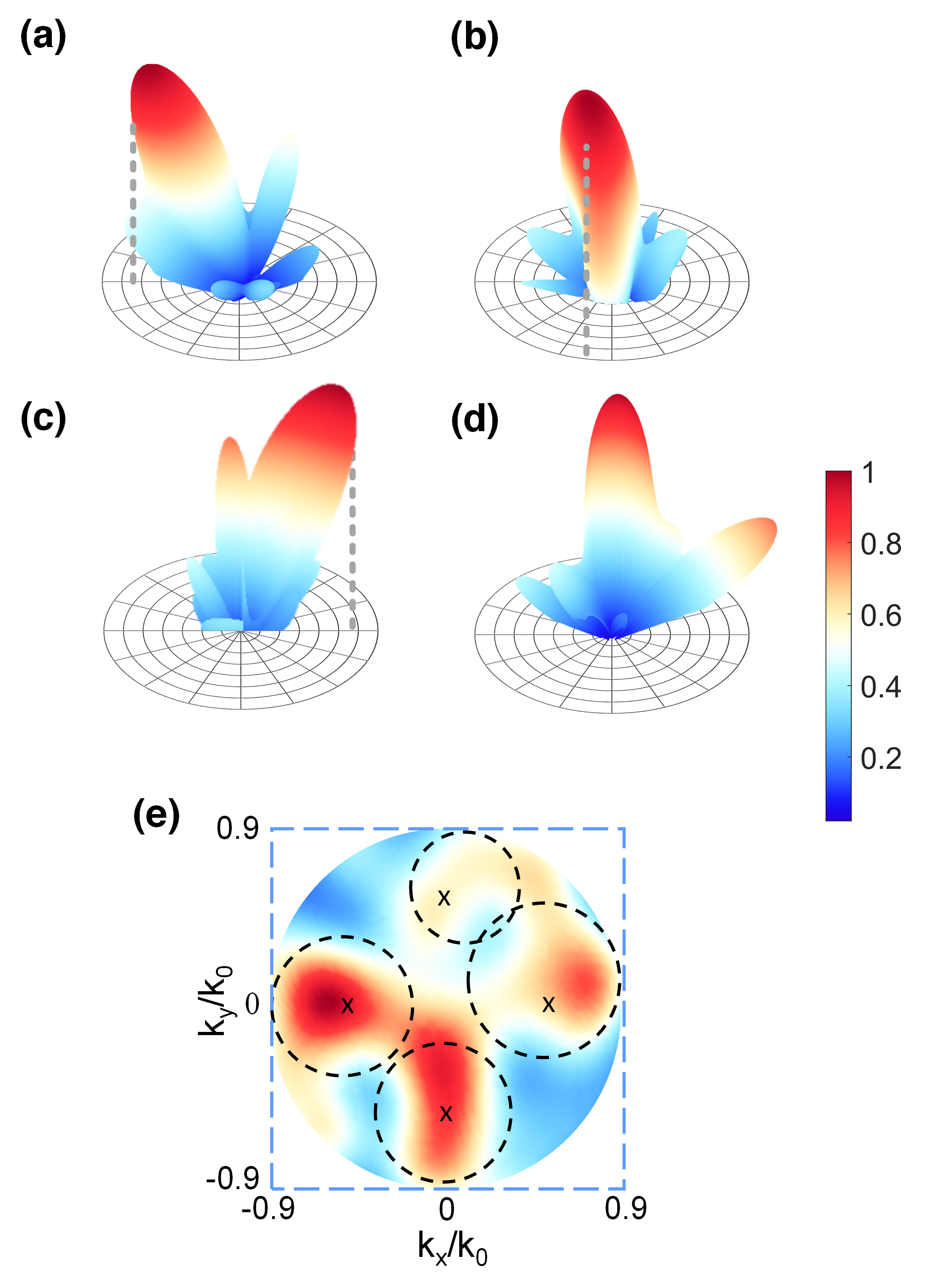}
	\caption{Measured far-field patterns. (a)-(d) Normalized 3-D patterns under the incidences of OAM modes $l_0=1, 2, -2, -1$; (e) Normalized 2-D pattern in $k$ space under the incidence of multiplexed OAM modes. The dotted black lines represent the half-power circles of far-field intensity patterns and the crosses represent the pre-designed scattered directions of the four OAM modes.}
	\label{farfield test}
\end{figure}
Measured far-field patterns are drawn in FIG.~\ref{farfield test}. The corresponding crosstalk matrix is
\begin{equation}
\mathbf{A_{ot}}=\left[\begin{array}{cccc}
1&0.30&0.17&0.29 \\
0.30& 1& 0.21& 0.11 \\
0.17& 0.21& 1& 0.33\\
0.29& 0.11& 0.33& 1\\
\end{array}\right].
\end{equation}
Compared with simulation results, the directions of the four scattered waves show some deviations, and there are also more severe side lobes, especially for modes -1 and -2. From the result of the multiplexed OAM incidences in 2-D $k$ space, mode crosstalks and wave deformations can be noticed, while the four modes can be roughly separated.  The measurement error can be largely attributed to the coupling between the edge of square hole and the upper surface of metasurface, the reflection between two metasurfaces, as well as the fabrication errors of metasurfaces. To realize the miniaturization of meta-atoms, working bandwidth has to be sacrificed to a relatively narrow level, while the high dielectric constant of substrate also imposes strict constraints for fabrication. Generally, the deviations in measurement are acceptable as our aim is to explore the feasibility of approaching the fundamental limit of OAM multiplexing in the practical design. Although our solution is proved to be reliable in numerical simulations and experiments, we believe more advanced meta-atoms should be proposed to realize wider working bandwidth and simpler fabrication process while maintaining small sizes, which will take full advantage of the wave manipulation ability of metasurfaces.
\section{Conclusion}
In summary, an accessible derivation of the upper scattering channel limit for OAM multiplexing is given by both angular-spectral analysis and rigorous EM Green's function method, which reveals the relation between the number of independent scattering channels and the physical size of detecting aperture. Then, the theory is verified with both ideal optimization case and practical full-wave simulation, where the minimax algorithm is applied successfully to suppress side lobes.  Moreover, microwave experiment is also carried out to test the performance of optimized metasurfaces.  By using this theory, the ultimate performance of SMM based communication system can be precisely estimated. The optimization algorithm is also useful for practical implementation of SMM, especially for OAM multiplexing.

\setcounter{equation}{0}
\setcounter{figure}{0}
\setcounter{subsection}{0}
\renewcommand\theequation{A\arabic{equation}}	
\renewcommand\thefigure{A\arabic{figure}}   
\section*{Appendix A: Connections between EM theory and information theory of a SMM system}
In view of information theory, SMM can be analyzed with the same model as the traditional MIMO system, as the underlying mathematics is universal. For a MIMO system, the ergodic information capacity per unit frequency is of the form \cite{capacity_MIMO}
\begin{equation}C=\mathcal{E}\left(\sum_{i=1}^{n} \log _{2}\left(1+\frac{P_i}{N} \sigma_{i}^{2}\right)\right),\end{equation}
where $\mathcal{E}$ is the mean operator, $n$ is the number of channels (i.e. DOF), $P_i$ is the signal power of each channel (usually equally allocated), $N$ is the power of noise and $\sigma_{i}^{2}$ are the squared singular values of the channel matrix. When the number of channels increases, the capacity will be enhanced with the same total input power, due to the fact that the DOF gain is more significant than the signal-to-noise ratio (SNR) gain embedded in the $\log$ operator, which is the spirit of MIMO technology.  It can be found that the capacity totally depends on the singular values.

In view of EM theory, the $\sigma_{i}^{2}$ in Eq. (A1) represent the eigenvalues of the EM eigenmodes in a SMM system \cite{piestun2000electromagnetic}, the values of $\sigma_{i}^{2}$ determine the strength of the corresponding eigenmodes. Each EM mode can be regarded as a channel, as well as the correlation between two channels is equivalent to the inner product between two EM modes. When the channels are uncorrelated with each other, i.e. the EM modes are independent and distinguishable, the maximum information capacity of the SMM system is approached.

\setcounter{equation}{0}
\setcounter{figure}{0}
\setcounter{subsection}{0}
\renewcommand\theequation{B\arabic{equation}}	
\renewcommand\thefigure{B\arabic{figure}}   
\section*{Appendix B: A rigorous deduction of the DOF limit between a 2-D plane and half space}
To clearly demonstrate the strict solution, let's begin with an intuitive method for deducing the communication DOF limit between two finite 2-D planes with arbitrary source distributions under paraxial approximation \cite{miller2000communicating}. For the source and receiving planes with areas $A_S$ and $A_R$, the minimum spot $a$ generated by source plane $A_S$ on receiver plane $A_R $ can be obtained from the concept of solid angle as
\begin{equation}
a \sim \lambda_0^2 D^2 / A_S,
\end{equation}
where $\lambda_0$ is the free space wavelength, $D$ is the distance between the two planes. As each spot can be utilized for independently coding information, the communication DOF can be intuitively regarded as the maximum containable number of spots on $A_R$
\begin{equation}
N \sim A_R / a,
\end{equation}
which is
\begin{equation}
N\sim\frac{A_{S} A_{R}}{\lambda_0^{2} D^{2}}.
\end{equation}
Similarly, if we replace the $A_R$ by a half space receiver (a circular area in angular spectrum defined by $k_x ^2+k_y ^2<k_0^2$), and replace $a$ by the minimum resolution area ($\Delta k_x\cdot \Delta k_y$) in angular spectrum, it will yield the intuitive method in Eq. \eqref{upper_limit}, this DOF refers to the number of significant EM modes in a communication system.

Now we can deduce the strict solution for the DOF limit.  First of all, for the communication between a source volume and a receiving volume, with arbitrary source distributions inside, it is easy to prove that the DOF limit is only dependent on the size of the volumes, and is independent of the specific source distributions, see Eqs. (3.3 - 4.2) in reference \cite{piestun2000electromagnetic}. What we need to do is to deduce the coupling operator $\mathbf{H}$ between a 2-D plane and half space, then to investigate the number of significant eigenmodes of this coupling operator. Without loss of generality, assume some point sources (delta function bases) are uniformly distributed on a 2-D plane, a similar method is utilized in deducing the coupling operator between two planes, see Eqs. (6 - 10) in reference \cite{miller2019waves}. The delta function basis can simplify the equations, and the results will not be affected by using other bases. As we consider an ideal far-field half space receiver, scalar Green's function is adequate for this scenario (two orthogonal polarization components are independent with each other). The scalar Green's function in free space is
\begin{equation}
g\left(\mathbf{r}, \mathbf{r}^{\prime}\right)=\frac{1}{4 \pi} \frac{\exp \left(-j k_0\left|\mathbf{r}-\mathbf{r}^{\prime}\right|\right)}{\left|\mathbf{r}-\mathbf{r}^{\prime}\right|},
\end{equation}
where $k_0$ is the free space wavenumber, $\mathbf{r}$ and $\mathbf{r^\prime}$ represent the positions of field and source. Under far-field approximation ($r \rightarrow \infty$), we have
\begin{equation}
\left|\mathbf{r}-\mathbf{r}^{\prime}\right| \approx	r-\mathbf{r}^{\prime} \cdot \mathbf{k}_{{R}},
\end{equation}
where $\mathbf{k}_{R}$ is the unit direction vector of $\mathbf{r}$, then the scalar Green's function becomes
\begin{equation}
g_{far}\left(\mathbf{k}_R, \mathbf{r}^{\prime}\right)=\frac{e^{-j k r}}{4 \pi r} \exp(j k_0 \mathbf{r}^{\prime}\cdot \mathbf{k}_R),
\end{equation}
which describes the field at $\mathbf{k}_R$ direction generated by a source point. As the term $(e^{-j k r}/4 \pi r)$ is a constant, it will not influence the DOF of this system. Considering a set of $N_S$ point sources at the positions $\mathbf{r}_{S n}$ in the source plane with the complex amplitudes $t_n$, and $N_R$ receiving directions at $\mathbf{k}_{Rm}$, the superposed electric field in direction $\mathbf{k}_{Rm}$ would be
\begin{equation}
E\left(\mathbf{k}_{Rm}\right)=\frac{e^{-j k r}}{4 \pi r} \sum_{n=1}^{N_{S}} \exp(j k_0 \mathbf{r}_{Sn}\cdot \mathbf{k}_{Rm}) t_{n}=\sum_{n=1}^{N_{S}} h_{mn} t_{n},
\end{equation}
where
\begin{equation}
h_{mn}=\frac{e^{-j k r}}{4 \pi r} \exp(j k_0 \mathbf{r}_{Sn}\cdot \mathbf{k}_{Rm})=g_{far}\left(\mathbf{k}_{Rm}, \mathbf{r}_{Sn}\right),
\end{equation}
\begin{figure}[ht!]
	\centering
	\includegraphics[width=3.4in]{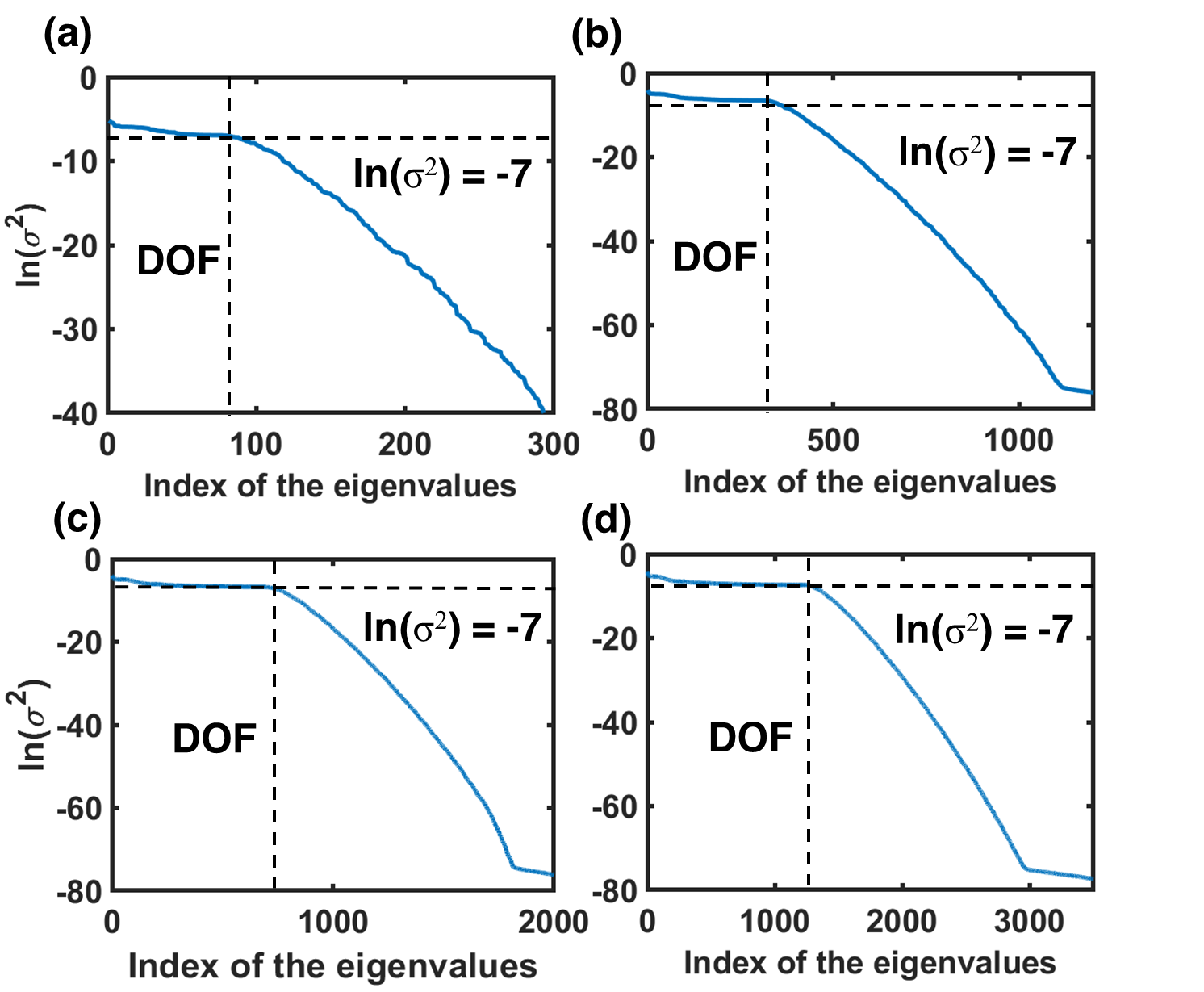}
	\caption{Eigenvalues of $\mathbf{H}\mathbf{H}^\dagger$, where $^\dagger$ denotes conjugate transpose. The intersection point of the two dot lines denotes the inflection point, the number of eigenvalues larger than the value of this inflection point denotes the number of dominant eigenmodes, i.e., DOF. (a) $L = 5\lambda_0$; (a) $L = 10\lambda_0$; (a) $L = 15\lambda_0$; (a) $L = 20\lambda_0$.}
	\label{MIMO}
\end{figure}
The received signals at one direction would be the sum of the fields from all the point sources added up at $\mathbf{k}_{Rm}$
\begin{equation}
f_{m}=\sum_{n=1}^{N_{S}} h_{mn} t_{n}.
\end{equation}
If $ t_{n}$ and $f_{m}$ are collected in the two column vectors ${t}=[t_1, t_2, \dots, t_{N_S}]^T$ and ${f}=[f_1, f_2, \dots, f_{N_R}]^T$, we can define the projection from the source plane to the half space as
\begin{equation}
{f}=\mathbf{H} {t},
\end{equation}
with
\begin{equation}
\mathbf{H}=\left[\begin{array}{cccc}
h_{11} & h_{12} & \cdots & h_{1 N_{S}} \\
h_{21} & h_{22} & \cdots & h_{2 N_{S}} \\
\vdots & \vdots & \ddots & \vdots \\
h_{N_{R} 1} & h_{N_{R} 2} & \cdots & h_{N_{R} N_{S}}
\end{array}\right],
\end{equation}
which is the coupling operator (channel matrix) between a source plane and half space. Then, with singular value decomposition (SVD), we can obtain the number of dominant eigenmodes of this system, which is the DOF limit. 
\begin{figure}[ht!]
	\centering
	\includegraphics[width=3in]{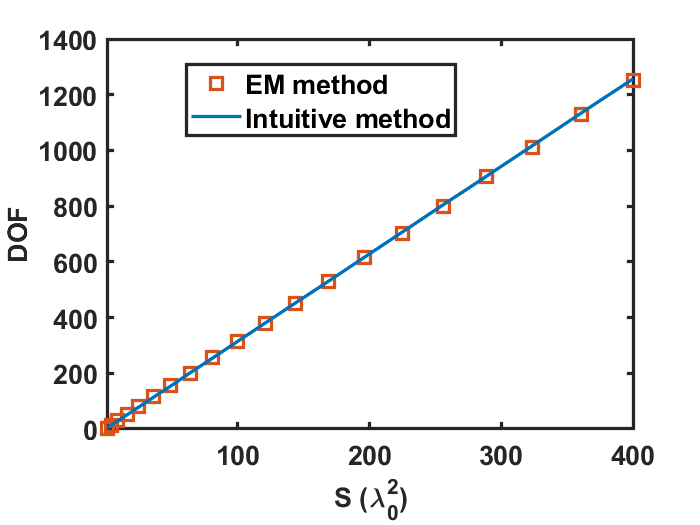}
	\caption{Relations between DOF and plane size calculated with the EM method and the intuitive method.}
	\label{MIMO}
\end{figure}

Assume $N_s\times N_s$ point sources are uniformly distributed in an $L\times L$ square plane, and the receiving $\mathbf{k}_R$ are the sampled directions in half space. Specifically, $\phi$ = 0 - 2$\pi$  is uniformly sampled with $4N_r$ points, and $\theta$ = 0 - $\pi$/2 is sampled with $N_r$ Gaussian-Legendre quadrature points as \cite{golub1969calculation}
\begin{equation}
\theta_{i} = \cos \left(\pi \frac{\frac{1}{2}+i}{N_r}\right)^+\cdot \frac{\pi}{2}
\end{equation}
where $i=[1,2,\cdots, N_r]$ and $^+$ denotes the positive part. To guarantee the information is totally captured by the ideal half space receiver, $N_r$ is set as $2L/\lambda_0$, which makes the results convergent. Now we can investigate the specific eigenvalues of the coupling operator. As depicted in FIG. B1, the eigenvalues of the coupling operators with different $L$ are demonstrated, these eigenvalues will decrease rapidly when pass the inflection point $\ln(\sigma^2) = -7$, thus the eigenvalues larger than $\ln(\sigma^2) = -7$ are regarded as the significant eigenvalues, and the DOF is the number of significant eigenvalues. Then, we can find the relation between DOF and the area of plane with this rigorous model, as depicted in FIG. B2. It is interesting to observe that the rigorous model gives almost the same result as the intuitive method.

Aforementioned results can be easily reproduced, as the delta function basis makes this procedure simplified. The reason for transforming OAM into plane-wave basis is that the plane-wave basis is naturally convenient for deducing the intuitive equation in angular spectrum, and is also particularly useful in microwave implementation. Therefore, the theoretical basis of the fundamental limit proposed in the manuscript is valid, and can be verified with both intuitive and rigorous methods.

\setcounter{equation}{0}
\setcounter{figure}{0}
\setcounter{subsection}{0}
\renewcommand\theequation{C\arabic{equation}}	

\section*{Appendix C: Minimax optimization}
The minimax algorithm is used to minimize the possible loss for a worst (maximum loss) scenario, which is frequently used in engineering problems \cite{madsen1978linearly, minimax}. The crucial part of constructing such algorithm is to set appropriate objective functions. The minimax optimization algorithm for our cases can be written as

\begin{equation}\label{minimax}
{\mathbf{minimax}}\{\mathbf{f}_n(\kappa_1,\kappa_2,\cdots\kappa_{100})\},\quad n=1, 2, 3, 4,
\end{equation}
where $\mathbf{f}_1,\,\mathbf{f}_2,\,\mathbf{f}_3,\,\mathbf{f}_4$ are the four objective functions under the incidences of four OAM modes and $[\kappa_1,\kappa_2,\cdots\kappa_{100}]$ denotes the input variables, being the phases of pixels $[\phi_1,\phi_2,\cdots\phi_{100}]$ for the ideal metasurface case and the rotation angles of meta-atoms $[\alpha_1,\alpha_2,\cdots \alpha_{100}]$ for the realistic metasurface case, respectively. With the input variables, four objective functions can be constructed with the normalized far-field patterns $P_{far}$ given in Section IV.

For the construction of objective functions, as each input OAM mode is related to one dash-line circle in angular spectrum (See Fig. 1b), the objective power is set as the sum of power included in the $k$ circle centered at a pre-designed direction. The corresponding wave numbers of the four critically designed directions are $a_n=k_{x}(n)/k_0=[0.65,0,-0.65,0]$ and $b_n=k_{y}(n)/k_0=[0,0.65,0,-0.65]$, where $k_0$ is the free-space wave number and $n=1,2,3,4$. The radius of the $k$ circle is set as $1/3$ $k_0$, which shows the best performance after testing. Note that the input powers of the four incident OAM modes are set to be the same, so that higher objective power indicates better performance. In order to search the best performance of the worst case under the incidences of four OAM modes, reciprocal of the objective power should be set as objective function. The final objective functions are
\begin{equation}\label{objective_function}
\mathbf{f}_n=\left[ \sum\limits_{(\frac{k_x}{k_0}-a_n)^2+(\frac{k_y}{k_0}-b_n)^2<(\frac{1}{3})^2}P_{far}(k_x,k_y)\right] ^{-1}.
\end{equation}

We use the fminimax function in MATLAB optimization toolbox to construct the algorithm.

\nocite{*}

\bibliography{apssamp}

\end{document}